# Interpreting Safety Outcomes: Waymo's Performance Evaluation in the Context of a Broader Determination of Safety Readiness

Francesca M. Favaro[a,1], Trent Victor[a], Henning Hohnhold[a], Scott Schnelle[a]

*[a] Waymo LLC, Mountain View, California*

**Abstract:** This paper frames recent publications from Waymo within the broader context of the safety readiness determination for an Automated Driving System (ADS). Starting from a brief overview of safety performance outcomes reported by Waymo (i.e., contact events experienced during fully autonomous operations), this paper highlights the need for a diversified approach to safety determination that complements the analysis of observed safety outcomes with other estimation techniques. Our discussion highlights: the presentation of a "credibility paradox" within the comparison between ADS crash data and human-derived baselines; the recognition of continuous confidence growth through in-use monitoring; and the need to supplement any aggregate statistical analysis with appropriate event-level reasoning.

**Introduction**

Waymo operates a fully autonomous commercial ride-hailing service - Waymo One - in Arizona and California. Waymo One showcases operations of the Waymo Driver - a SAE Level 4 ADS that is, by definition, responsible for the entirety of the Dynamic Driving Task execution without reliance on human intervention.

The safety of Waymo's on-road operations relies on rigorous approaches for safety readiness determination: Waymo's Safety Readiness Methodologies, first presented by Webb et al. (2020), have been refined over 10+ years of operations, and leverage a combination of simulation testing, closed-track testing, and public road testing (Waymo, 2021).

Recently, we released data from our first one million miles driven in Rider-Only (RO) configuration (i.e., without an autonomous specialist on board overseeing the ADS) (Victor et al., 2023). Out of 20 contact events experienced by the Waymo Driver, the data shows that there were no reported injuries, and that only two collisions were comparable to those reported in the National Highway Traffic Safety Administration's Crash Investigation Sampling System ("CISS") - a nationally representative database of collisions that includes police reported events in which at least one vehicle was towed.

This is not the first time Waymo released performance data about the Waymo Driver. In October 2020, we released data from over six million miles of autonomous driving (Schwall et al., 2020). The majority of those had a trained autonomous specialist on-board overseeing the Waymo Driver operations - what we call an Autonomous Driving Configuration (ADC). There are several differences between the performance data reported in 2023 and that in 2020, including:

— *The sourcing mileage*. Data leveraged in the analysis comes from distinct types of operations of the Waymo Driver: RO operations for the 2023 paper vs. on-road testing data in autonomous mode with an autonomous specialist on board for the 2020 paper;

— *Actual vs. simulated collisions*. While all of the 20 contact events included in the 2023 paper stem from observed events on public roads, the 2020 paper also included what we term "counterfactual" collision events, or "what-if" simulations. Those are situations in which we predicted a contact event through an analysis of post-disengagement simulation, thus predicting what would have likely happened had the autonomous specialist not regained control of the vehicle during on-road testing. The 2020 dataset consisted of 47 contact events, divided across 18 actual and 29 simulated contacts, none of which would be expected to result in severe or life-threatening injuries (Schwall et al., 2020).

It is important to understand these differences and their role in informing the determination of safety readiness for an ADS. In this short paper, we share our perspective on collection, interpretation, and confidence estimation in relation to ADS performance outcomes, and frame such insights within the broader context of Waymo's safety determination practices, as published in (Favaro et al., 2023).

---

[1]Corresponding author: fmfavaro@waymo.com

**Discussion**

One of the most prominent uses of ADS performance data is to enable the comparison with the current state of transportation safety, to ensure that the introduction of this technology could in fact fulfil the sought potential for a positive safety impact. In fact, the comparison of an ADS collision rate[2] with that of the current transportation ecosystem[3] can support the determination of having achieved an adequate level of safety to field the technology (see (Favaro, 2021) and discussion therein). The practical undertaking of such comparison presents, however, complex challenges that require the utmost care and engineering rigour to approach, as we note below.

The analysis of safety-outcomes (i.e., collision events) undertaken to inform readiness determination through a comparison with baselines derived from human data (e.g., current crash statistics from public databases) entails calibrating for a number of limitations that impact both the data observed during operations of the ADS as well as of the data employed to generate an appropriate baseline for comparison. The first consideration, impacting both the ADS as well as the baseline, could be termed a *"credibility paradox"* and is notionally represented in Figure 1.

**Figure 1.** Notional representation of the credibility paradox

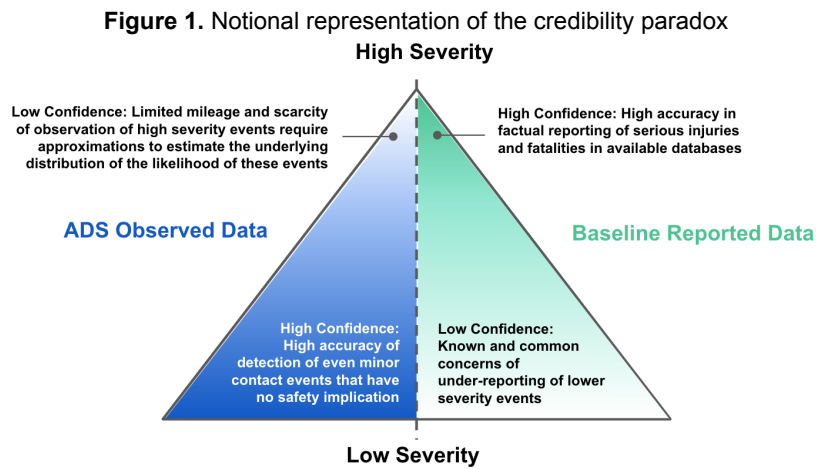

*Source:* Own elaboration

As observed for both the data released by Waymo in 2023 (Victor et al.) and in 2020 (Schwall et al.), the vast majority of ADS events result in no injury (and, in many situations, also no property damage). Detection of even very minor events results in high statistical confidence for low severity outcomes for the ADS. In contrast data employed in the generation of low severity baselines is plagued by well-known concerns of under-reporting for minor events, where an estimated 24.3% of injury-bearing collisions and 59.7% of property-damage-only collisions end up not reported to police in NHTSA's CISS collision database (Blincoe et al., 2015).[4] Conversely, for high severity events, accurate reporting of serious injuries crashes is available in most countries, but evaluation of performant ADS systems will, even at the large testing scale of million of miles that Waymo implements, have limited statistical power due to the low number of high severity events. The credibility paradox thus cautions us against simplistic approaches to undertake this type of comparisons, which, while useful, can be concerningly misguided if not adequately approached. To address these difficulties, Waymo adopts a number of strategies, such as:

— We augment, when informative and available, publicly available crash data with large-scale naturalistic data and/or dash-cam data from private providers. This can help improve statistical confidence in the baseline for low severity events.

---

[2] A rate can be obtained by normalising collision counts by, for example, hours of operation or mileage driven.
[3] Or a subset thereof, such as human-driven vehicles of the same category.
[4] Under-reporting of events with no property damage is actually expected to be much higher than that for property-damage-only events, even though the literature does not provide a precise estimate. These types of no property damage events are even rarer to find in databases used for the computation of baselines, further skewing the contrast with ADS minor event rates.

- We seek and analyse in-depth those ADS low severity events that may have resulted in more serious consequences under different situations.[5] This concept, referred to sometimes as "mining the diamond" (Smith and Jones, 2013), can be enabled by metrics that quantify the potential for higher injury risk, like, for example, the Maximum Injury Potential in Kusano and Victor (2022).
- We assess confidence in the comparison between ADS performance and the baseline across multiple categories of event types and severity potential. On one hand, the usage of an organised typology of conflicts for horizontal categorization of events (applied to both the ADS data and the baseline data) can better pinpoint lower/higher statistical confidence across certain types of interactions (e.g., lower reported data for the baseline associated with rear-end collisions), which helps guide data sourcing efforts listed in the prior bullet. On the other hand, systematically breaking out collision events by injury risk ("severity") for vertical categorization of events enables us to identify plausible performance trends despite the known data limitations, which we appropriately weight within the broader determination of safety.
- We leverage, within our readiness determination, a collection of metrics and testing approaches to provide coverage of situations that may carry higher severity potential than what was encountered or observed on the road. An example is Waymo's Collision Avoidance Testing (see Kusano et al., 2022) where the Waymo Driver's performance is assessed in situations requiring immediate emergency action on the part of the ADS.

The evaluation of readiness of a new SW release pre-deployment (which is, by definition, a prediction) through the analysis of collision rates thus combines a number of sources, including: i) actual observations of RO deployments from prior SW versions; ii) observations from public testing in ADC; iii) simulations that may involve fully synthetic testing and/or hybrid counterfactual simulation of data partially collected on the road. Reliance on one type of data versus the other may also change over time, depending on maturity and scale of the operations of a company. This is represented in Figure 2, where we also visually showcase differences amongst the 2020 vs. the 2023 datasets. Reliance on data sourced from ADC operations and simulation is expected before deployment. As previously explained, any comparison to baselines should be appropriately contextualised and complemented by other methodologies to establish RO readiness for public road operations. In-use monitoring of the Waymo Driver's performance enables continuous confidence growth post RO deployment.

**Figure 2.** Change in reliance on diverse data sources with deployment stage - trends for illustrative purposes only

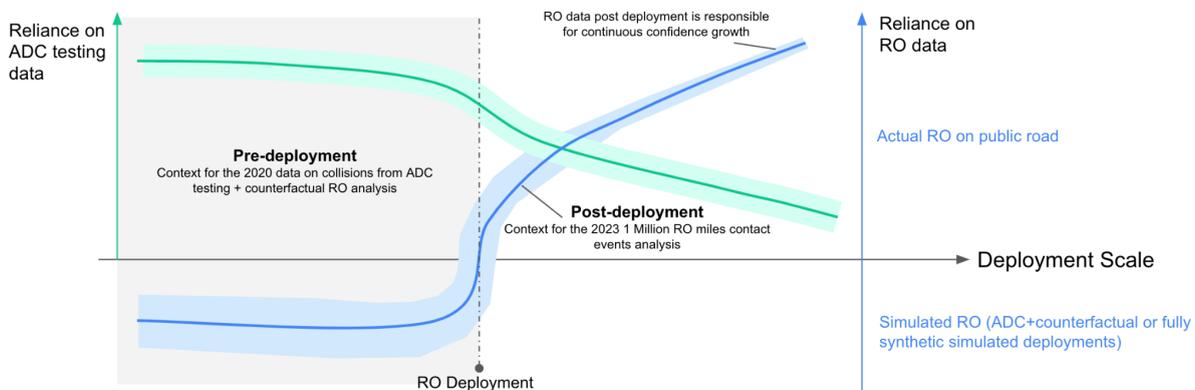

*Source:* Own elaboration

Additionally, the usage of safety performance outcomes for the determination of safety can lead to over-indexing on aggregate performance indicators that inadvertently conceal the presence of undesirable levels of risk in individual events or scenarios. This is conveyed in Figure 3, where Waymo's approach to safety calls for a balance between event-level acceptance criteria, which sample risk attributable to individual instances of occurrence and support event-level risk assessment, and aggregate-level acceptance criteria, which work as overarching indicators of performance and are not necessarily traceable back to individual events (Favaro et al., 2023).[6]

---

[5] For example, information such as the fact that none of the RO events reported in our one million miles paper were intersection-related or involved Vulnerable Road Users can allow us to better establish the potential for high severity outcomes.
[6] The possible lack of traceability between individual events and aggregate rates is due to potential estimation processes (e.g., extrapolation) that make the evaluation of ADS behaviour in each event infeasible.

The assertion that the Waymo Driver is successful at reducing injuries and fatalities is thus grounded in analyses that go beyond the prediction of fatality rates. In (Victor et al., 2023) we point to how an appropriate comparison with human baseline could be made (i.e., by ensuring compatibility of events with standardised reporting requirements, such as those in CISS, and by looking at both the ability to reduce frequency of events and/or mitigate severity of outcomes). Still, other research studies and methodologies at Waymo help ensure that the level of residual risk for Waymo's deployed fleet is acceptable. The analysis in (Scanlon et al., 2022), for example, pointed to appropriate conflict avoidance performance of the Waymo Driver, showcasing the ability to avoid entering a conflict in the first place and showcasing an additional mechanism for crash outcomes prevention.

**Figure 3.** Considerations on unintentional over-indexing on aggregate performance indicators

*Source:* Favaro et al., 2023

## Conclusions

Confidence in the determination of safety of the Waymo Driver increases continuously, as both credibility of available data and the validity of our predictions improves over time thanks to in-use monitoring of on-road data. Understanding of reliance on different types of data and/or methodologies, paired with the appropriate understanding of limitations within non-ADS data employed for the generation of baselines, is an important starting point to ensure evaluation of a developer claim of safety can be appropriately contextualised. The combination of Waymo's safety methodologies provides a balanced and responsible approach to confidently evaluate performance of the Waymo Driver within the broader context of our safety determination lifecycle (Victor et al., 2023; Schwall et al., 2020, Kusano et al., 2022; Scanlon et al., 2022; Favaro et al., 2023).